\documentclass[12pt]{article}
\usepackage{amssymb,amsmath,epsf,cite,times}

\numberwithin{equation}{section}

\newcommand{\vz}{\vec\zeta}

\newcommand{\bu}{{\bf u}}
\newcommand{\ba}{{\bf a}} 
\newcommand{\bai}{{\bf a}_{int}}

\newcommand{\be}{{\bf e}}

\newcommand{\bk}{{\bf k}}
\newcommand{\bn}{{\bf n}}
 
\newcommand{\br}{{\bf r}}
\newcommand{\bdr}{{\bf \delta r}}

\newcommand{\bB}{{\bf B}}

\newcommand{\bL}{ \hat {\bf L}}
\newcommand{\cL}{{\cal L}}

\newcommand{\hL}{{\hat L}}
\newcommand{\hH}{{\hat H}}
\newcommand{\hHi}{{\hat H}_{int}}

\newcommand{\OO}{\Omega_0} \newcommand{\OOO}{\Omega_0^{\,2}}
\newcommand{\bOO}{\overline{\Omega}_0}

\newcommand{\Ob}{\Omega_b}
\newcommand{\Obb}{\Omega_b^{\,2}}

\newcommand{\Os}{\Omega_\bot}

\newcommand{\e}{\epsilon}
\newcommand{\ee}{\epsilon^{\,2}}

\newcommand{\ddz}[1]{\langle  z^2( #1)\rangle}
\newcommand{\ddx}[1]{\langle  x^2( #1)\rangle}
\newcommand{\ddy}[1]{\langle  y^2( #1)\rangle}

\newcommand{\lla}{\langle\!\langle}
\newcommand{\rra}{\rangle\!\rangle}

\newcommand{\la}{\left\langle}
\newcommand{\ra}{\right\rangle}
\newcommand{\lb}{\left( }
\newcommand{\rb}{\right) }

\newcommand{\Kbb}{K_b^{\,2}}

\newcommand{\vtt}{v_{th}^{\,2}}
\newcommand{\vt}{v_{th}}

\newcommand{\beq}{\begin{equation}}
\newcommand{\eeq}{\end{equation}}

\newcommand{\bea}{\begin{eqnarray}}
\newcommand{\eea}{\end{eqnarray}}

\renewcommand{\Re}{{\rm \bf Re}}
\renewcommand{\Im}{{\rm \bf Im}}

\newcommand{\rhoo}{\rho_0^{\,2}}

\begin{document}


\begin{center}
  
  {\large Diffusion in a stochastic magnetic field.}

\vspace{0.5cm}

\parbox[c]{13.5cm}{{\small We consider a stochastic differential
    equation for a charged particle in a stochastic magnetic field,
    known as A-Langevin equation. The solution of the equation is
    found, and the velocity correlation function $\langle u(t)^i\,
    u(0)^j\rangle$ is calculated in Corrsin approximation. A
    corresponding diffusion constant is estimated. We observe
    different transport regimes, such as quasilinear- or Bohm-type
    diffusion, depending on the parameters of plasma. }}

  \vspace{0.5cm}
  
  {\small D.~Lesnik, S.~Gordienko, M.~Neuer, K.-H.~Spatschek \\
    \today }

\end{center}

\vspace{0.5cm}

The problem of transport in a magnetic field is in focus of a number
of theoretical and experimental works performed during the last
several decades. The reason for such extraordinary interest to this
problem is caused by the problem of still unsuccessful plasma
confinement. A very strong deviation of the diffusion rate from naive
classical predictions is due to the nonlinear effects, which have to
be taken into account.  On the other hand, the character of transport
phenomena is often very counterintuitive and paradoxical, what
cautions one to be especially careful with any theoretical
predictions.

In our work, we use the stochastic differential equations approach to
the problem of transport of charged particles in a magnetic field
\cite{VanCampen}.  There are two traditional ways of considering this
problem. The first one usually is referred to as V-Langevin equations.
In this approach one considers stochastic equations for a guiding
center of a test particle in a drift approximation
\cite{RaxWhite1992,Balescu1994,AllTogether1995,BalescuEijnden1996}.

We concentrated on the second approach named A-Langevin equations,
that invokes an exact equation of motion of a single test particle,
for which interaction forces are mimicked by phenomenological damping
and acceleration terms. On the base of the solution of the equation of
motion one can calculate a velocity correlation function that leads to
the diffusion tensor. Generally the exact solution of the problem is
not possible, or at least, extremely complicated.  Nevertheless it is
still possible to make some estimations in different limiting cases
assuming that the perturbation of the magnetic field is weak.

The paper is organized as follows. In the first section we give the
mathematical formulation of the problem and present the solution of
the equation of motion as well as the velocity correlation function,
expressed through the Lagrange correlator of the magnetic field. In
the second part we estimate the Lagrange magnetic field correlator in
the Corrsin approximation. This allows us to reformulate the problem
in terms of a differential equation for the mean square displacement.
In the third part we find some particular solutions of this equation
and present the corresponding diffusion constants. All mathematical
details of the derivations are placed in appendices.

\section{Equation of motion and formulation of the problem}

The equation of motion of a charged particle with the charge $Z e$ in
a magnetic field $\bB(t)$, that experiences damping and random
acceleration ($-\nu\,\bu(t)$ and $\ba(t)$ respectively) is given by:
\begin{equation}
  \label{eq:a1}
  {\bf \dot u}(t) =
  \frac{Z\,e}{mc}\,[\bu(t) \times \bB(t)] -\nu\,\bu(t) +\ba(t)\,.
\end{equation}
where $\nu$ is an effective collisions frequency.

This is a stochastic differential equation, known as A-Langevin
equation. There are two stochastic functions in this equation: the
stochastic magnetic field $\bB(t)$ and the random acceleration
$\ba(t)$. To close the system of equations, the stochastic properties
of these functions should be defined. We assume that both are Gaussian
processes, which means that the first and the second order correlation
functions provide a complete statistical description of those
functions. Concerning the magnetic field it means that we consider
only the regions of plasma with completely chaotic magnetic field, not
containing structures like KAM surfaces or islands.

For the random acceleration $\ba(t)$ we chose the white noise
approximation, i.e.
\begin{eqnarray}
  \label{eq:acor}
  \langle a(t)^i\rangle=0\,, \quad \langle a(t_1)^i\,a(t_2)^j\rangle = 
  A\,\delta^{ij}\,\delta(t_1-t_2)\,. 
\end{eqnarray}
It is well known from the theory of Brownian particle that the
equilibrium thermal velocity $v_{th}$ is related to the collisions
frequency $\nu$ and value $A$ as $\vtt= A/2\,\nu$\,. This relation
remains valid for charged particles in a magnetic field as well, since
equilibrium thermal velocity is not affected by the Lorenz force.


The magnetic field is supposed to consist of two parts: a constant
component $B_0$ directed along the $z$-axis and a small perturbation
with zero average:
\[
\bB(\br,t)= B_0\,{\bf e}_z + 
\e\,B_0\,\vz(\br,t)\,;\qquad {\rm where}\quad
\langle |\vz|^2\rangle=1\,,\quad
\langle \vz \,\rangle= 0\,.
\]
The perturbation parameter $\e$ is supposed to be small.  The {\em
  Eulerian} correlation function (ECF) of the magnetic field
\begin{equation}
  \label{ECF}
  {\cal E}(|\bdr|,|\delta t|)=
  \langle\zeta(\br_1,t_1)^i\,\zeta(\br_2,t_2)^j\rangle \,,\qquad
  \bdr=\br_1-\br_2\,,\quad \delta t= t_1-t_2\,, 
\end{equation}
will be specified in subsequent sections.  As we will see later, the
ECF appears in the calculation in implicit form. The {\em Lagrange}
correlation function (LCF)
\begin{equation}
  \label{LCF}
  {\cal L}^{ij}(|\delta t|)=
  \langle\zeta(\br_1(t_1),t_1)^i\,\zeta(\br_2(t_2),t_2)^j\rangle 
\end{equation}
appears explicitly instead. In contrast to the ECF, which is
calculated in fixed points of space, the LCF is taken in points of the
stochastic trajectory, and thus cannot be obtained from the stochastic
properties of the magnetic field solely. To perform the calculation,
we have to make a hypothesis of statistical independence of the LCF
from other stochastic processes.  Within the framework of this
hypothesis one can perform averaging over all stochastic processes
independently from one another. After the averaging, the mean
displacement $\langle \br^2(t)\rangle$ as well as the velocity
correlation function will be found from the self consistency
condition.

The object of our interest is the ``running'' diffusion coefficient,
that is defined in the usual way as
\[D_i(t)= \frac{1}{2}\,\frac{d}{dt}\,\langle r_i^{\,2}(t)\,\rangle\,;
\qquad i= x,\, y,\, z\,.
\]
The mean square displacement (MSD) $\langle r_i^{\,2}(t)\,\rangle$ can
be derived in turn from the velocity correlation matrix:
\begin{eqnarray}
  \label{eq:MSDmatrix}
  \la r^i(t)\,r^j(t) \ra = \! \int_0^t\!\!\!\int_0^t \! dt_1\,dt_2\,
  \la u^i(t_1)\,u^j(t_2) \ra = 
  2\!\int_0^t \! (t-\tau)\, \la u^i(\tau)\,u^j(0) \ra \,d\tau \,.
\end{eqnarray}
Combining the last two formulas, we write the diffusion tensor as
\begin{eqnarray}
  \label{eq:dx}
  D_{ij}(t)=\int_0^t \!\! \la u^i(\tau)\,u^j(0) \ra \, d\tau\,.
\end{eqnarray}
As one sees, the problem is now reduced to the calculation of the
velocity correlator $\la u^i(\tau)\,u^j(0) \ra$ in the Lagrange frame
of reference.

The solution of the Eq.~(\ref{eq:a1}) can be found straightforwardly.
First we transform this equation into standard form, introducing the
generators of the $SO(3)$ group $\bL = (\hat L_i)_{jk} = -
\epsilon_{ijk}$\,. The corresponding finite rotation matrices are
$R_i(t)=\exp(t\,\hL_i)$. We define the ``magnetic field operator'' as
\begin{equation}
  \label{eq:H}
  \hH(t)= \vz(\br(t),t)\cdot \bL= \zeta_1(\br(t),t)\,\hL_1+
  \zeta_2(\br(t),t)\,\hL_2+\zeta_3(\br(t),t)\,\hL_3\,,
\end{equation}
where $\br(t)$ is the particle trajectory. Introducing the 
unperturbed part of Lamor frequency $\OO= Z\,e\,B_0/(mc)$,
we can now rewrite Eq.~(\ref{eq:a1}) as
\begin{equation}
  \label{eq:a10}
    {\bf \dot u}(t) = [-\nu -\OO\,\hL_3 -\e\,\OO\,\hH(t)]\,\bu(t)
    +\ba(t)\,.
\end{equation}
In terms of the Green function the solution is
\begin{equation}
  \label{eq:solution}
  \bu(t)= e^{-\nu\,t}\,R_3(-\OO\,t)\,
  \Big\{
  G(t)\,\bu_0+\int_0^t G(t',t)\,\bai(t')\,dt'
  \Big\}\,,
\end{equation}
where the propagator $G(t_2,t_1)$ is defined as:
\begin{eqnarray}
  \label{eq:nb2}
  G(t_2,t_1)= T\,\exp
  \left(-\e\,\OO \int_{t_2}^{t_1} \hHi(\tau)\,d\tau
    \right) \,; \qquad  G(t):=G(0,t)\,,
\end{eqnarray}
and $T\,\exp(\,)$ denotes a time-ordered exponent. The values
$\hHi(t)$ and $\bai(t)$ are magnetic field operator and random
acceleration in interaction representation:
\begin{eqnarray}
\label{eq:Hint}
\hHi(t)=  R_3(\OO\,t)\,\hH(t)\, R_3(-\OO\,t) \,; \quad
\bai(t)  =  e^{\nu\,t}\,R_3(\OO\,t)\,\ba(t)\,.
\end{eqnarray}

Now we can calculate the velocity correlation function. On the base of
the solution~(\ref{eq:solution}) we construct a product
$u^i(t_1)\,u^j(t_2)$, that should be averaged over all the stochastic
functions -- random acceleration $\ba(t)$, initial velocities $\bu_0$
and magnetic field fluctuations $\vz(t)$.  The expression for the
product $u^i(t_1)\,u^j(t_2)$ becomes rather bulky though
straightforward, so we relegate the calculation of the velocity
correlator to the appendix~\ref{VCF}\footnote{\label{factorize} Here
  we only emphasize again that the averaging procedure is made under
  the assumption of statistical independence of LCF and other
  stochastic processes. This assumption allows us to factorize the
  averaging of the velocity correlator in order to get an explicit
  expression in terms of LCF and MSD. This factorization procedure is
  analogous to the Corrsin approximation, and is reliable for a small
  coupling parameter~$\e$.}. After averaging, the velocity correlation
function becomes a reduced matrix with diagonal elements:
\begin{subequations}
  \label{eq:vcf} 
  \begin{align}
    \label{eq:uuxy}
    \langle u(t_1)^i\,u(t_2)^i\,\rangle =& \, 
    \Re\left[ \vtt
      e^{-\nu\tau-\gamma(\tau)} e^{-i\,\bOO\,\tau}
    \right]
    +\ee\vtt \cL_x(\tau)\, e^{-\nu\tau-\gamma(\tau)},\quad & &i= 1,2  \\
    \label{eq:uuz}
    \langle u(t_1)^i\,u(t_2)^i\,\rangle =&\, \vtt \,e^{-\nu\,\tau}\,
    +2\,\ee\,\vtt \,
    e^{-\nu\,\tau}\,\cL_x(\tau)\,\cos\,\bOO\tau\,, \hspace{2cm}& &i=3    
  \end{align}
\end{subequations}
where $\tau=|t_1-t_2|$,\, $\vt$ is the equilibrium thermal velocity,
$\bOO=\OO+\ee\,\OO$ is the renormalized Lamor frequency, $\cL_x(t),
\cL_y(t), \cL_z(t)$ are the diagonal elements of the LCF, and
\begin{equation}
\label{eq:gamma(t)_1}
\gamma(t)=\ee\,\OOO\!
\int_0^t \! (t-\tau) \cL_z(\tau)\, d\tau= \ee\,\OOO\! \int_0^t \!
d\tau \int_0^\tau \!d\tau'\, \cL_z(\tau')\,.
\end{equation}

Further analysis will be possible after having found an equation for
the LCF. But here some general remarks can be made about the
expectations for the diffusion constant.

For the subsequent calculation, we suppose that the magnetic field is
strong enough, so that the time $\tau_0=1/\OO$ is the shortest
characteristic time of the problem. Due to the fast oscillating
multiple $e^{-i\,\bOO\,\tau}$ the first term of the correlator
(\ref{eq:uuxy}) does not give a significant contribution to the
diffusion coefficient.  A detailed analysis shows (see below) that the
first term leads to the classical diffusion coefficient in an
unperturbed magnetic field $D_{cl}=\vtt\,\nu/\OOO$. An effect of
magnetic field fluctuations is thus contained in the second term.  The
rate of decay of the correlation function is an outcome of two
different mechanisms of decorrelation. They are collisions (due to the
term $\nu\,t$ in the exponent) and field fluctuations (due to the
function $\gamma(t)$ in the exponent).  As one can see from the
Eq.~(\ref{eq:dx}), this rate is crucial for the diffusion coefficient.

We also note the fact that the velocity correlator (\ref{eq:uuxy}) is
almost the same as the one for guiding centers (see App.~\ref{sec:GC},
Eq.~\ref{eq:fR}), but, in contrast to the latter, it has an additional
exponential multiple $e^{-\gamma(t)}$. This multiple leads to
essential differences between the diffusion coefficient of particles
and that of guiding centers.

From here we use $\OO$ instead of\, $\bOO$ for brevity.

\section{Lagrange correlation function of magnetic field}
\label{sec:Ansatz}

\newcommand{\ls}{\lambda_\bot}
\newcommand{\lss}{\lambda_\bot^{\,2}}
\newcommand{\lp}{\lambda_\|}
\newcommand{\lpp}{\lambda_\|^{\,2}}

The right choice of a magnetic field correlation function ${\cal
  E}^{ij}(|\br_1-\br_2|)=\la \zeta^i(\br_1)\,\zeta^j(\br_2)\ra$ is
still an open question. Many authors who investigated neoclassical
transport in magnetized plasma used the correlator in a Gaussian form.
In the general case such a correlator can be presented as
\[
{\cal E}^{ij}(\br)=\la \zeta(\br)^i\,\zeta(0)^j\ra=
C^{\,ij}(\br)\,
\exp\left( 
  -\frac{x^2}{\lss}-\frac{y^2}{\lss}-\frac{z^2}{\lpp} 
\right);\quad i=1,\, 2,\,3\,;
\]
where the tensor function $C^{\,ij}(\br)$ should be chosen to
make the correlator ${\cal E}^{ij}(\delta \br)$ satisfy the
divergence-free condition. Detailed analysis of this problem can be
found for example in \cite{AllTogether1995}, where a correlator for a
two-dimensional field perturbation was proposed. In this paper it is
shown that, although the Euler correlator ${\cal E}^{ij}(\delta \br)$
has off-diagonal elements, the Lagrange correlation tensor (measured
in a frame of reference moving with a magnetic field line) is a
diagonal matrix.

One possible choice for the correlation matrix, satisfying the
divergence-free condition, is
\begin{eqnarray}
  \label{eq:divfreetensor}
  {\cal E}^{ij}(\br)= \left( 
    \frac{\partial^2}{\partial r^i\,\partial r^j}-
    \delta^{ij}\,\frac{\partial^2}{\partial r^k\,\partial r^k}
  \right)\,C(\br)
\end{eqnarray}
with arbitrary function $C(\br)$. It corresponds to the correlation
function of the vector potential $\langle A(\br)^i\,A(0)^j
\rangle = C(\br)\,\delta^{ij}$. Choosing
$C(\br)=\exp(-x^2/\lss-y^2/\lss-z^2/\lpp)$ and omitting ``small''
terms, we come to the following form of the ECF
\begin{eqnarray}
  \label{eq:ECF}
  {\cal E}^{ij}(\br)= \la \zeta(\br)^i\,\zeta(0)^j \ra = 
  \left(
    \begin{array}[c]{ccc}
      1 & 0 & 0 \\
      0 & 1 & 0 \\
      0 & 0 & \varsigma^2
    \end{array}
  \right)\,
  \exp
  \left(
    -\frac{x^2}{\lss}-\frac{y^2}{\lss}-\frac{z^2}{\lpp} 
  \right); \quad \br=\{x,y,z\}\,;
\end{eqnarray}
which we choose as a starting hypothesis for the subsequent
calculations. Here $\varsigma^2= \frac{2\,\lpp}{\lss+\lpp}$.

Notice that in the limit $\ls\gg\lp$ the $z$-component of the
magnetic field perturbation and the $zz$-component of ECF vanish. In
this case we approach a degenerated 2-dimensional problem.

We use the Corrsin approximation \cite{Corrsin,McComb} to estimate the
Lagrange correlator, given the Eulerian one:
\begin{eqnarray}
  \label{eq:corrsin}
  \cL(t)^{ij}=\la \zeta(t)^i\,\zeta(0)^j\ra &=
  & \int \langle\,\zeta(\br)^i\,\zeta(0)^j \,
  \delta(\br- \br(t))\rangle\,d^3 r \nonumber \\
  &\approx & \int\langle\,\zeta(\br)^i\,\zeta(0)^j \rangle\,\langle
  \delta(\br- \br(t))\rangle\,d^3 r\,.
\end{eqnarray}
Here $\cL(t)^{ij}$ is the Lagrange correlator of the magnetic field, $\la
\zeta(\br)^i\,\zeta(0)^j\ra$ -- the Eulerian correlator,
$\la\delta(\br-\br(t))\ra$ is the averaged particle propagator.
Treating $\br(t)$ as a stochastic variable, we apply the cumulant
expansion to the Fourier representation of the $\delta$-function to
obtain
\begin{eqnarray}
  \label{eq:qe2}
  \la \delta(\br-\br(t))\ra= 
  \frac{\exp
    \left(
      - \frac{x^2}{2\,\ddx{t}} 
      - \frac{y^2}{2\,\ddy{t}}
      - \frac{z^2}{2\,\ddz{t}} 
    \right)}
  {(2\,\pi)^{3/2}\, \sqrt{\ddx{t}\,\ddy{t}\,\ddz{t}}} \,.
\end{eqnarray}
Here we used the assumption that the trajectory $\br(t)$ is a Gaussian
process, i.e. the only nontrivial cumulants of the displacement
$\br(t)$ are: $\lla x^2(t)\rra$, $\lla y^2(t)\rra$ and $\lla
z^2(t)\rra$.

Now substituting (\ref{eq:qe2}) into (\ref{eq:corrsin}) and performing
the integration, we can express the diagonal elements of the Lagrange
correlator in terms of the MSD:
\begin{eqnarray}
  \label{eq:LCFx}
  &&\cL_x(t)= \cL_y(t)=\frac{1}
  {
    \left( 1+ \Gamma_x(t) / \lss  \right)
    \sqrt{ \left( 1+ \Gamma_z(t) / \lpp  \right)}
  }\,, \\
  \label{eq:LCFz}
  &&\cL_z(t)= \varsigma^2\,\cL_x(t);
\end{eqnarray}
where we have designated 
\(
\Gamma_x(t)= \lla x^2(t) \rra\,, \ 
\Gamma_y(t)= \lla y^2(t) \rra\,, \ 
\Gamma_z(t)= \lla z^2(t) \rra\,,
\)
and $\varsigma$ is defined after Eq.~(\ref{eq:ECF}).

Next, taking the second time derivative of the
Eq.~(\ref{eq:MSDmatrix}), we obtain the equations for the MSD:
\begin{align}
  \label{eq:Gammax}
  &\Gamma_{\perp}''(t)=
  2\,\Re \left[ \vtt \,e^{-\nu\,t-\gamma(t)}\, 
  e^{-i\,\OO\,t}\right]+  2\,\ee\,\vtt \,\cL_x(t)\,
  e^{-\nu\,t-\gamma(t)}\,, \\
  \label{eq:Gammaz}
  &\Gamma_z''(t)= 2\,\vtt \,e^{-\nu\,t}\,
  +4\,\ee\,\vtt \,
  e^{-\nu\,t}\,\cL_x(t)\,\cos\,\OO t\,,
\end{align}
where $\Gamma_\perp(t)=\Gamma_x(t)=\Gamma_y(t)$. These equations
should be solved with initial conditions \mbox{$\Gamma_i(0)=0$},
$\,\,\Gamma_i'(0)= 0$\,. The function $\gamma(t)$ is defined in
Eq.~(\ref{eq:gamma(t)_1}).

The system of equations (\ref{eq:LCFx}), (\ref{eq:LCFz}),
(\ref{eq:Gammax}) and (\ref{eq:Gammaz}) completes the self consistent
description of the MSD as well as LCF.  One can solve this system with
respect to the LCF and obtain an integral equation for the LCF, which
stands on its own interest. An analogous procedure was performed by
Hai-Da~Wang {\it et al.} \cite{AllTogether1995}, where an integral equation was
obtained for the LCF of a magnetic field line.

Another possibility is to exclude $\cL(t)$ to obtain an equation for
the MSD. This equation will lie in the focus of our interest in the
next section.

\section{Estimation of the MSD and diffusion coefficient}

Let us first consider the motion in $z$-direction. From the equation
(\ref{eq:Gammaz}) one can see that the second term in the
right-hand-side is $\ee$-small with respect to the first one and thus
can be neglected. The equation becomes uncoupled and can be
integrated, giving
\begin{eqnarray}
  \label{eq:GammazSolution}
  \Gamma_z(t)= \frac{2\,\vtt }{\nu^2}\,
  \left(\nu\,t+e^{-\nu\,t}-1\right)=
  \left\{
    \begin{array}[c]{ll}
      \vtt\,t^2\,, \qquad \qquad & \nu\,t \ll 1\,; \\
      (2\,\vtt/\nu) \,t\,,       & \nu\,t \gg 1\,.
    \end{array}
  \right.
\end{eqnarray}
We see that the effect of the magnetic field fluctuations on the
motion along the magnetic field appears as a small correction (of
order less then $\ee$) to the pure collisional
result~(\ref{eq:GammazSolution}).

To analyze the perpendicular motion, we split equation
(\ref{eq:Gammax}) into two parts:
\begin{subequations}
  \begin{align}
    \label{eq:G1}
    &\Gamma_x(t)= \Gamma_{x1}(t)+\Gamma_{x2}(t)\,; \\
    \label{eq:Gx1}
    &\Gamma_{x1}''(t)= 2\,\Re \,\vtt \,e^{-\nu\,\tau-\gamma(\tau)}\, 
    e^{-i\,\OO\,\tau}\,,     &&   
    \Gamma_{x1}(0)= 0\,,\ \Gamma_{x1}'(0)= 0\,; \\
    \label{eq:Gx2}
    &\Gamma_{x2}''(t)= 2\,\ee\,\vtt \,\cL_x(\tau)\,
    e^{-\nu\,\tau-\gamma(\tau)}\,, && 
    \Gamma_{x2}(0)= 0\,,\ \Gamma_{x2}'(0)= 0\,;
  \end{align}
\end{subequations}

\subsection{Estimation of $\Gamma_{x1}(t)$.}

Consideration of the first part $\Gamma_{x1}(t)$ is simplified due to
the fast oscillating multiple~$e^{-i\,\OO\,t}$ in the right-hand-side
of Eq.~(\ref{eq:Gx1}). Performing a Fourier transform and noting that
the spectrum of the function in the right-hand-side of the equation
has a sharp maximum at $\omega= \OO$, we expand the spectrum around
this point. After inverse transform the equation can be integrated,
giving 
\begin{equation}
  \label{eq:Gammax1}
  \Gamma_{x1}(t)= 2\,\rhoo\,\nu\,t+
  2\,\rhoo\,
  \left(
    1-\cos(\OO\,t)\,e^{-\nu\,t-\gamma(t)}
  \right)\,,
\end{equation}
where $\rho_0=v_{th}/\OO$ is the Lamor radius.

\subsection{Estimation of $\Gamma_{x2}(t)$}

\newcommand{\tG}{\tilde\Gamma_{x2}}

To perform the analysis of $\Gamma_{x2}(t)$ it is useful to introduce
the so-called {\em decorrelation time} $t^*$, that is defined as a
solution of the equation $\nu\,t^*+\gamma(t^*)=1\,$. For the rough
estimation of the solution of the Eq.~(\ref{eq:Gx2}) we replace
$e^{-\nu\,\tau-\gamma(\tau)}\approx\theta(t^*-\tau)$\,.  Thus for
times $t<t^*$ the solution of the exact equation (\ref{eq:Gx2}) can be
approximated by the solution of the subsidiary equation
\begin{subequations}
  \label{eq:eq1fortGamma} 
  \begin{eqnarray}
    \label{eq:equationfortGamma}
    && \tG''(t)= 2\,\ee\,\vtt \,\cL_x(t)=
    2\,\ee\,\vtt
    \left( 
      1+\frac{\Gamma_{x1}(t)+\tilde\Gamma_{x2}(t)} {\lss}  
    \right)^{-1}
    \left( 1+\frac{\vtt} {\lpp}\,t^2  \right)^{-1/2} \,; \\
    &&\tG(0)= 0\,,\ \tG'(0)= 0\,; 
  \end{eqnarray}
\end{subequations}
and for $t>t^*$ the solution is a linear function of time:
\begin{eqnarray}
  \label{eq:GammafromtGamma}
  \Gamma_{x2}(t)\approx 
  \left\{
    \begin{array}[c]{ll}
      \tG(t)\,,\quad & t\leq t^*\,; \\
      \tG(t^*)+ \tG'(t^*)\,(t-t^*)\,,\qquad& t>t^*\,;
    \end{array}
  \right.  
\end{eqnarray}
As soon as the solution of the equation (\ref{eq:eq1fortGamma}) is
found, the decorrelation time $t^*$ can be obtained from the equation
\begin{equation}
  \label{eq:dectime}
  \nu\,t^*+ \frac{\varsigma^2}{2\,\rhoo}\,\tG(t^*)=1\,.
\end{equation}
Indeed, the definition (\ref{eq:gamma(t)_1}) of the function
$\gamma(t)$ yields
\begin{equation}
  \label{eq:gamma(t)_2}
  \gamma(t)= \frac{\varsigma^2}{2\,\rhoo}\,\tG(t)\,,
\end{equation}
which gives rise to Eq.~(\ref{eq:dectime}). The diffusion
coefficient follows immediately from~(\ref{eq:Gammax1})
and~(\ref{eq:GammafromtGamma})\,:
\begin{equation}
  \label{eq:DiffCoeff}
  D_x=\frac{1}{2}\,\frac{d}{dt}\,\Gamma_x(t\to\infty)= 
  \frac{1}{2}\,\Gamma'_{x1}(t\to\infty)+
  \frac{1}{2}\,\tG'(t^*)=
  \frac{\vtt\,\nu}{\OOO}+\frac{1}{2}\,\tG'(t^*)\,.
\end{equation}
Below we find some particular solutions of 
Eq.~(\ref{eq:eq1fortGamma}) and the corresponding diffusion
coefficient in different limiting cases that are characterized 
by the ratio of the correlation lengths $\lp$ and $\ls$.

\subsubsection{The case $\ls \to \infty$}
\label{subsec:ls}
As it was already mentioned, in this limiting case we have
$\varsigma\to 0$, and the $\cL_z(t)$ component of the LCF vanishes. As
a result, the velocity decorrelation, caused by the magnetic field
fluctuations, becomes very slow due to the multiplier $\varsigma$
appearing in the definition (\ref{eq:gamma(t)_1}) of the function
$\gamma(t)$.

Eq.~(\ref{eq:equationfortGamma}) can be rewritten in the
form:
\begin{eqnarray}
  \label{eq:Gamma_inf}
  && \tG''(t)= 
  2\,\ee\,\vtt
  \left( 1+ \Obb\,t^2  \right)^{-1/2},
\end{eqnarray}
giving rise to the solution
\begin{eqnarray}
  \label{eq:sldfkj}
  &&\tG(t)= 2\,\ee\,\lpp\,
  \left(
    1+ \Ob\,t\,{\rm ArcSinh}(\Ob\,t) - \sqrt{1+\Obb\,t^2}
  \right)\,,
\end{eqnarray}
where we have introduced $\Ob= v_{th}/\lp\,,{\ \ } K_b=
\e\,\varsigma\,\OO/\Ob$. The $\tG(t)$ has the following asymptotic
scaling:
\begin{eqnarray}
  &&\tG(t) \approx 
  \left\{
    \begin{array}[c]{ll}
      \ee\,\vtt\,t^2\,,   & \text{for \ \ }  t\ll 1/\Ob\,;\\
      2\,\ee\,\lpp\,\Ob\,
      t\,\ln(\Ob\,t)\,, \quad         & \text{for \ \ }  t\gg 1/\Ob\,;
    \end{array}
  \right.
\end{eqnarray}
The dimensionless parameter $K_b$ is the Kubo number, which
characterizes the influence of the stochastic magnetic field on the
averaged motion of a particle. Quantitatively it measures the
contribution of magnetic fluctuations to the decorrelation of the
particle velocity, and hence to the diffusion constant. To
describe the influence of the collisions on the diffusion we
introduce a dimensionless parameter \mbox{$K_\nu=\nu/\Ob$.}
Considering different limiting cases of the parameters $K_\nu$ and
$K_b$, we obtain the following transport regimes:
\begin{enumerate}
\item First we consider the case of weak perturbation $\Kbb\ll K_\nu
  \ll 1$.  The diffusion coefficient coincides up to a logarithmic
  correction factor with the well-known quasilinear result
  $D_{ql}\sim\vt\,\ee\,\lp$:
  \begin{eqnarray}
    \label{eq:Dx2}
    D_x \approx  
    \frac{\vtt\,\nu}{\OOO}+
    \vt\,\ee\,\lp \,\ln(\Ob/\nu)\,.
  \end{eqnarray} 
  The decorrelation time is $t^*\approx 1/\nu$\,.
\item In the case $K_\nu\ll \Kbb\ll 1$ the transport regime is also of
  quasilinear type with a slightly different correction multiplier\,:
  \begin{eqnarray}
    \label{eq:Dx3}
    &&D_x \approx \frac{\vtt\,\nu}{\OOO}+
    \vt\,\ee\,\lp \,\ln(\alpha)\,,\quad
    \mbox{where}\quad \alpha= 
    \frac{1}{\Kbb\,\ln\left(1/\Kbb \right)}\,,
  \end{eqnarray}
  and the decorrelation time becomes $t^*=\alpha/\Ob$.
\item The strong collisional case corresponds to the choice $K_\nu\gg
  \text{max}\left(1,\,K_b\right)$. The corresponding decorrelation
  time is $t^*\approx 1/\nu$ and the diffusion constant
  \begin{eqnarray}
    \label{eq:Dx1}
    D_x \approx \frac{\vtt\,\nu}{\OOO}+ \frac{\ee\,\vtt}{\nu}\,.
  \end{eqnarray}
\item The last case $K_b\gg\text{max}\left(1,\,K_\nu \right)$
  considers strong magnetic fluctuations. The decorrelation time is
  $t^*\approx 1/(\e\,\OO\,\varsigma)$. In this case we observe a
  Bohm-type diffusion with characteristic scaling $D\sim B^{-1}$:
  \begin{eqnarray}
    \label{eq:Dx4}
    D_x \approx  \frac{\vtt\,\nu}{\OOO}+
    \frac{\e\,\vtt}{\varsigma\,\OO}\,.
  \end{eqnarray}
\end{enumerate}
Now we can specify more carefully the condition $\ls\to\infty$. 
To be able to neglect the terms in the first bracket in the
right-hand-side of the Eq.~(\ref{eq:equationfortGamma}), the $\ls$
should satisfy the condition:
\begin{eqnarray}
  \Gamma_x(t^*)/\lss \ll \text{min} 
  \left(\vtt\,(t^*)^2/\lpp \,,\, 1\right)\,,
\end{eqnarray}
where for rough estimation we can take $\Gamma_x(t^*)\approx D_x \,
t^*$\,.

\subsubsection{The case $\lp\to\infty$}
\label{subsec:lp}

Similarly to the previous case, the condition $\lp\to\infty$ can
be written as
\begin{eqnarray}
  \vtt\,(t^*)^2/\lpp \ll \text{min} 
  \left(\Gamma_x(t^*)/\lss\,,\,1\right),
\end{eqnarray}
where we substitute $\Gamma_x(t^*)\approx D_x \, t^*$, and the
corresponding diffusion coefficients are calculated below.

The Eq.~(\ref{eq:eq1fortGamma})
in the considered limit becomes
\begin{subequations}
  \begin{eqnarray}
    \label{eq:tGammax22}
    && \tG''(t)= 
    2\,\ee\,\vtt
    \left( 
      1+\frac{\Gamma_{x1}(t)+\tilde\Gamma_{x2}(t)} {\lss}  
    \right)^{-1}\,; \\
    &&\tG(0)= 0\,,\ \tG'(0)= 0\,; 
  \end{eqnarray}
\end{subequations}
where the exact expression for $\Gamma_{x1}(t)$ is represented in the
Eq.~(\ref{eq:Gammax1}). Since we are only interested in the solutions
of the equation above for times $t<t^*$, we can approximately
substitute \(\Gamma_{x1}(t)\approx 2\,\rhoo (1-\cos(\OO\,t))\).  The
obtained equation has in its right-hand-side a slowly varying function
of time with superimposed fast oscillations of small amplitude due to
the term $(\rhoo/\lss) \cos(\OO\,t)$. So we can average the equation
over the fast oscillations and obtain a simplified equation describing
the slow dynamics of the $\tG(t)$. Performing the integration over the
period of oscillations we can treat $\tG(t)$ as constant during the
integration time.  Making change of variables $U(t)=U_0+\tG(t)/(2\,\rhoo)$,
we come to the following autonomous equation:
\begin{eqnarray}
  \label{eq:U}
  &&U''(t)= \frac{a^2}{2\,\sqrt{U^2(t)-1}}\,;  \\
  &&U'(0)= 0\,;\quad U(0)= U_0\,; \nonumber
\end{eqnarray}
where $a= \e\,\vt\,\ls /\rhoo ,{\ \ }U_0=1+\lss/(2\,\rhoo)$.  This
equation can be integrated analytically:
\begin{eqnarray}
  \label{eq:WWW}
  \frac{\sqrt{\pi}}{2}\,
  \left( b\,{\rm Erfi}(W(t))-\frac{1}{b}\,{\rm Erf}(W(t))\right)=
  a\,t\,;
\end{eqnarray}
where $b=U_0+\sqrt{U_0^{\,2}-1}$\,,
Erf($t$) and Erfi($t$) are error functions, and $\tG(t)$ can be
expressed through $W(t)$ as
\begin{eqnarray}
  \label{eq:GGG}
  \tG(t)= \rhoo\,\left( b\,e^{W^2(t)}+\frac{1}{b}\,e^{-W^2(t)} \right).
\end{eqnarray}
The solution for $\tG(t)$, given by the
Eqs.~(\ref{eq:WWW},\,\,\ref{eq:GGG}), can now be
used for the calculation of the decorrelation time $t^*$ and the
diffusion constant in different limiting cases. One readily finds an
explicit expression for asymptotic behavior of $\tG(t)$:
\[\tG(t)=\left\{
  \begin{array}[c]{ll}
    \ee\,\vtt\,\kappa^2\,t^2 \quad    & \Os\,t\ll 1 \\
    2\,\e\,\vt\,\ls\,t\,\sqrt{\ln( \alpha\,\Os t)}\qquad   
    & \Os\,t\gg 1 
  \end{array}
\right.
\]
where we have introduced the effective perpendicular magnetic
frequency
\[
\Os=\frac{\e\,v_{th}}{\ls}\, \kappa^2\,,\qquad
\kappa^2=\frac{\ls}{\sqrt{\lss+ 4\,\rhoo}}\,,
\quad \,\alpha= \frac{8\,\kappa^2}{(1+\kappa^2)^2}.
\]

\newcommand{\Ks}{K_{\bot\,b}} 
\newcommand{\Kss}{K_{\bot\,b}^{\,2}}
\newcommand{\Kn}{K_{\bot\,\nu}}

Similarly to the previous section, we estimate the decorrelation time
$t^*$ from Eq.~(\ref{eq:dectime}), where we substitute
$\varsigma=\sqrt{2}$ in the limit considered, and find the diffusion
constant from Eq.~(\ref{eq:DiffCoeff}).  There are four different
asymptotic diffusion regimes, classified by the dimensionless
parameters $\Ks= \e\,\OO\,\kappa/ \Os$ and $\Kn= \nu/\Os$:
\begin{enumerate}
\item The case of weak perturbation $\Kss\ll \Kn\ll 1$ leads to a
  Kadomtsev-Pogutse-type diffusion $D_{KP}\sim\vt\,\e\,\ls$ with a
  logarithmic correction. In this case we find $t^*\approx 1/\nu$\,,
  and the diffusion coefficient
  \begin{eqnarray}
    \label{eq:Dxs2}
    &&D_x \approx  \frac{\vtt\,\nu}{\OOO}+
    \vt\,\e\,\ls\,\sqrt{\ln(\alpha\,\Os/\nu)}\,.
  \end{eqnarray}
\item In the case of \mbox{$\Kn\ll \Kss\ll 1$} the obtained result is
  also of Kadomtsev-Pogutse-type, but with different correction
  multiplier\,:
  \begin{eqnarray}
    \label{eq:Dxs3}
    &&D_x \approx \frac{\vtt\,\nu}{\OOO}+
    \vt\,\e\,\ls\,\sqrt{\ln(\beta)}\,,\quad \text{where}\quad \beta= 
    \frac{\alpha}{\Kss\,\sqrt{\ln( \alpha/\Kss )}}\,.
  \end{eqnarray}
  The decorrelation time in this case is 
  $t^*\approx 1/(\Kss\,\Os\,\ln(\alpha/\Kss))$.
\item Strong collisional diffusion corresponds to the choice
  $\Kn\gg\text{max}(1,\,\Ks)$. In this case we find $t^*\approx 1/\nu$
  and the diffusion coefficient:
  \begin{eqnarray}
    \label{eq:Dxs1}
    D_x \approx \frac{\vtt\,\nu}{\OOO}+ \frac{\ee\,\vtt\,\kappa^2}{\nu}\,.
  \end{eqnarray}
\item The case $\Ks\gg \mbox{max}(1,\,\Kn )$ deals with strong
  magnetic fluctuations. We find a Bohm-type diffusion with
  characteristic scaling $D\sim B^{-1}$. The diffusion coefficient is
  \begin{eqnarray}
    \label{eq:Dxs4}
    D_x \approx  \frac{\vtt\,\nu}{\OOO}+
    \frac{\e\,\vtt\,\kappa}{\OO}\,,
  \end{eqnarray}
  and the decorrelation time $t^*$ is given by $t^*\approx
  1/(\e\,\OO\,\kappa)$.
\end{enumerate}

\section{Discussion}

We have investigated some aspects of transport of charged particles in
a stochastic magnetic field. On the base of the formal solution of a
stochastic differential equation (A-Langevin equation) we obtained the
velocity correlation function (\ref{eq:vcf}). It is expressed through
the correlation functions of stochastic processes effecting the
trajectory, such as stochastic magnetic field or random acceleration.
Next we found an equation for the Lagrange correlation function of the
magnetic field (Eq.~(\ref{eq:LCFx})) in Corrsin approximation. This
equation closes the system of equations for the determination of the
mean square displacement as a function of time. The diffusion constant
follows then as a time derivative of the MSD.
  
Since the equations obtained cannot be integrated in general case, we
have considered only limiting cases with respect to the correlation
lengths $\ls$ and $\lp$. The results are presented in sections
\ref{subsec:ls} and \ref{subsec:lp}.
  
A structure of the diffusion constant becomes more clear from the
following simple consideration. We see from Eq.~(\ref{eq:vcf}) that
the velocity correlator is a exponentially decreasing function of time
with amplitude $\ee\,\vtt$. We conclude that the diffusion coefficient
(\ref{eq:dx}) can be roughly estimated as $D\sim \ee\,\vtt\,t^*$,
where $t^*$ is the decorrelation time of the velocity correlator (the
width of exponent).  Hence it is the decorrelation time, what
basically defines the transport rate.
  
The two mechanisms of the velocity decorrelation -- magnetic field
fluctuations and collisions -- are characterized by the two
dimensionless parameters (Kubo numbers): $K_b$ and $K_\nu$ in the
limiting case $\ls\to\infty$, and $\Ks$ and $\Kn$ in the case
$\lp\to\infty$. The transport regime (i.e., scaling of the diffusion
coefficient) appears to be dependent solely on the ratio of these
parameters.
  
We have classified the obtained results by the ratio of the Kubo
numbers. In the case of ``weak'' perturbation, when both Kubo numbers
are less then 1, we have found the well-known results: the quasilinear
regime in the limit $\ls\to\infty$, and the Kadomtsev-Pogutse result
in the limit $\lp\to\infty$.  In the case of ``strong'' perturbation
the collisional or Bohm-like diffusion has been found. We note that
the Rechester-Rosenbluth result is more complicated to obtain
analytically, since it cannot be calculated in the limiting cases
$\ls\to\infty$ or $\lp\to\infty$.

It is interesting to note that the limiting case
$K_b\gg\text{max}(1,\,K_\nu)$ includes also a theoretical limit
$\OO\to\infty$. In this case we find the diffusion constant vanish,
yet the guiding center diffusion does not (see the note at the end of
the App.~\ref{sec:GC}). Indeed, the stronger the magnetic field is,
the better the approximation holds true that a single particle
describes a spiral path along a magnetic field line, i.e. the plasma
is frozen in a strong magnetic field. As long as we keep the
perturbation parameter $\e$ constant, the field lines keep diffusing,
and so do the guiding centers of particles. The spurious contradiction
is due to the well-known phenomenon, that in a magnetized plasma the
transport of guiding centers differs from the transport of particles
themselves \cite{Braginskii}. We see that there exists only a region
of plasma parameters, beyond which the guiding center approximation
fails to describe the transport processes in magnetized plasma.

\appendix

\section{Derivation of the velocity correlation function}
\label{VCF}

The velocity correlation function (VCF) can be obtained from the exact
solution~(\ref{eq:solution}). For this purpose we construct a product
of two components of the velocity $u(t_1)^i\,u(t_2)^j$ and then
average over the stochastic processes. As it was mentioned in the main
text (see the footnote on the p.~\pageref{factorize}), it is not
possible to perform the averaging over different stochastic processes
independently, since the magnetic field $\vec\zeta(\br(t))$, appearing
in the solution~(\ref{eq:solution}), is an implicit function of all
other stochastic values. However, if the magnetic fluctuations are
small, we can make the hypothesis that the Lagrange correlator of the
magnetic field is statistically independent of the other stochastic
values -- initial velocity $\bu_0$ and random acceleration~$\ba(t)$.
This hypothesis allows the factorization of the velocity correlation
function, somewhat analogous to the well-known Corrsin approximation.
 
First we average the product $u(t_1)^i\,u(t_2)^j$ over the initial
velocities $\bu_0$ and random accelerations~$\ba(t)$. For the
averaging, we use the following assumption about initial velocity
distribution:
\begin{eqnarray}
  \label{eq:nb4}
  \langle u_0^{\,i} \rangle= 0;\quad 
  \langle u_0^{\,i}\, u_0^{\,j}\rangle = v_{th}^2 \, \delta^{ij} \,.
\end{eqnarray}
A calculation leads to the following expression for the correlation
matrix in index notation:
\begin{equation}
  \label{eq:g1}
  \la u(t_1)^i\,u(t_2)^j\ra_{u_0,\,a(t)} = 
  \vtt \,\,e^{-\nu\,(t_1-t_2)}\,
  R_3\Bigl(-\OO\,t_1\Bigr)^i_k \,
  G(t_2,t_1)^k_l\,R_3\Bigl(\OO\,t_2\Bigr)^l_j \,,
\end{equation}
where $R_3(t)=\exp(t\,L_3)$ is a finite rotation matrix, and the
propagator $G(t_2,t_1)$ is defined in the Eq.~(\ref{eq:nb2}). To
obtain this result, the following properties of the matrix
$G(t_2,t_1)$ were used:
\[G(t)^T=G(t)^{-1}\,,\qquad G(t',t_1)\,G(t',t_2)^T=G(t_2,t_1)\,.\]
Both equations follow from the general properties of a time-ordered
exponent and from the fact that $G(t_2,t_1)$ is an exponent of a skew
symmetric matrix.

Next the equation (\ref{eq:g1}) should be averaged over the stochastic
magnetic field $\vz(\br(t),t)$, which appears in the exponent of the
propagator $G(t_2,t_1)$. To calculate the average of the exponent of a
stochastic variable, we need to apply the cumulant expansion. The
generalization of the cumulant expansion \cite{VanCampen} on the
time-ordered exponent is
\begin{eqnarray}
  \label{eq:cumexp}
  \lefteqn{\la T\,\exp\lb \int_0^t \! V(\tau)\,d\tau \rb \ra=}\\
  && \exp \lb 
  \int_0^t dt_1\,\la V(t_1)\ra +
  \frac{1}{2!}\int\!\!\int_0^t dt_1\,dt_2\, 
  T \lla V(t_1)\,V(t_2)\,\rra + \cdots \rb =  \nonumber \\
  &&  \exp \lb 
  \int_0^t dt_1\,\la V(t_1)\ra +
  \int_0^t \!dt_1 \!\! \int_0^{t_1}\! dt_2\, 
  \lla V(t_1)\,V(t_2)\,\rra + \cdots \rb , \nonumber 
\end{eqnarray}
where $T \lla\,\cdot\, \rra:= \lla T\,\cdot\, \rra $ is a time-ordered
cumulant. Now we apply this formula to the Eq.~(\ref{eq:g1}),
substituting \( V(t)= -\e\,\OO\,\hHi(t)\). Making use of the
definitions~(\ref{LCF}), (\ref{eq:H}) and (\ref{eq:Hint}),
we obtain
\begin{eqnarray}
  \label{eq:hi2}
  \langle G(t_2,t_1) \rangle = \langle G(\tau) \rangle =\exp 
  \left\{ -\ee\,\OOO \int_0^\tau \!\! (\tau-\tau')\,A(\tau')\, d\tau'
  \right\}\,,
\end{eqnarray}
where
\[A(\tau')=    \left(
      \begin{array}[c]{ccc}
        \cL_x(\tau')\,\cos(\OO\,\tau')+\cL_z(\tau')&-\cL_x(\tau')\,\sin(\OO\,\tau')&0\\
        \cL_x(\tau')\,\sin(\OO\,\tau') & \cL_x(\tau')\,\cos(\OO\,\tau')+\cL_z(\tau')&0\\
        0               & 0        & 2\,\cL_x(\tau')\,\cos(\OO\,\tau')
      \end{array}
    \right)\,,\]
and $\tau= |t_1-t_2|$. The matrix in the exponent has a reduced
form, and hence we have two invariant subspaces: $\{x,y\}$ and
$\{z\}$. In complex number representation (see
App.~\ref{app:xy_plane}) we replace in the $\{x,y\}$ aggregate:
\[
\hat L_3 \to i\,;\qquad (\hat L_3)^2\to -1\,; 
\qquad \hat R_3(\alpha)\to e^{i\,\alpha}\,;
\]
Thus, the propagator can now be rewritten in the form:
\begin{eqnarray}
  \label{eq:hu12} 
  \lefteqn{\langle G(\tau)\rangle= }\\
  &&\exp
  \left\{
    -\ee\,\OOO\!
    \int_0^\tau \! (\tau-\tau')
    \left(
      \begin{array}[c]{cc}
        \Big(\cL_x(\tau')\, e^{i\,\OO\,\tau'}+\cL_z(\tau')\,\ \Big) &
        \begin{array}[c]{c}
          0\\
          0
        \end{array}\\
        0    \qquad  \quad   0  & \ 2\,\cL_x(\tau')\,\cos(\OO\,\tau')
      \end{array}
    \right)\,d\tau'\,
  \right\}.  \nonumber
\end{eqnarray}
The diagonal elements of the $\{x,y\}$ aggregate of the propagator can
be obtained as the real part of the complex valued expression above.

Note that, after averaging, the matrix $R_3$ commutes with $\la
G(\tau) \ra$, and thus the whole velocity correlator becomes:
\begin{eqnarray}
  \label{eq:hu14}
  \langle u(t_1)^i\,u(t_2)^j \rangle =\la u(\tau)^i\,u(0)^j \ra=
  v_{th}^2\,e^{-\nu\,\tau}\,R_3(-\OO\,\tau)\,\la G(\tau)\ra\,;\quad
  \tau= |t_1-t_2|\,;
\end{eqnarray}

Now we can write the results of integration for $\{x,y\}$ and $\{z\}$
subspaces independently. For diagonal elements we get:
\begin{eqnarray}
  \label{eq:hu19}
  \langle u(t)\,u(0)\,\rangle = 
  \left\{
    \begin{array}[c]{ll}
      \Re  \left[ \vtt \,e^{-\nu\,t}\,e^{-i\,\OO\,t}\,
      e^{-\gamma(t)}\,
      e^{-\mu(t)}\right];\qquad       & \text{for}\quad \{x,y\}\\
      \vtt \,e^{-\nu\,t}\,
      e^{-2\,\Re [\mu(t)]};   & \text{for}\quad \{z\}\,.
    \end{array}
  \right.
\end{eqnarray}
Here we have used the designations:
\begin{eqnarray}
  \label{eq:hu20.1}
  &&\gamma(t)=\ee\,\OOO\!
  \int_0^t \! (t-\tau)
  \cL_z(\tau)\, d\tau=
  \ee\,\OOO\! \int_0^t \! d\tau  \int_0^\tau \!d\tau'\,
  \cL_z(\tau')\, ;\\
  \label{eq:hu20}
  &&\mu(t)=\ee\,\OOO\!
  \int_0^t \! (t-\tau)
  \cL_x(\tau)\, e^{i\,\OO\,\tau}\,d\tau=
  \ee\,\OOO\!
  \int_0^t \! d\tau \! \int_0^\tau \! d\tau'\,
  \cL_x(\tau')\, e^{i\,\OO\,\tau'}\,.
\end{eqnarray}
The formula for the velocity correlator can be simplified if we use
the reasonable assumption that the Lamor frequency $\OO$ defines the
shortest characteristic time in the system (limit of strong magnetic
field). Noting that the Fourier spectra of the integrated function has
a sharp maximum at $\omega =\OO$, we come to the asymptotic (as
$\OO\to\infty$) formula for $\mu(t)$:
\begin{eqnarray}
  \label{eq:hu33}
  \mu(\tau)& = &\ee\,i\,\OO\,\tau+ 
  \ee\, \Big( 1 -\cL_x(\tau)\,e^{i\,\OO\,\tau}\Big)\,.
\end{eqnarray}
The first term in this expression has, in fact, the form of a secular
term of a perturbation expansion. It can be nullified by the
renormalization of the frequency $\OO$.
Designating $\bOO=\OO+\ee\,\OO$\, and omitting the non-relevant terms
of the order $\ee$, we can write for the $\{x,y\}$ component of the
correlator:
\begin{eqnarray}
  \label{eq:hu35.1}
  \langle u(t_1)\,u(t_2)\,\rangle =\Re \,
  \vtt \,e^{-\nu\,\tau-\gamma(\tau)}\, 
  e^{-i\,\bOO\,\tau}+  \ee\,\vtt \,\cL_x(\tau)\,
  e^{-\nu\,\tau-\gamma(\tau)}\,.
\end{eqnarray}

In an analogous way we obtain an expression for the $\{z\}$ component.
Without loss of accuracy we can replace $\OO$ by $\bOO$ and omit
non-relevant corrections of the order $\ee$, finally coming to
\begin{eqnarray}
  \label{eq:hu36.3}
  \langle u(t_1)\,u(t_2)\,\rangle =\vtt \,e^{-\nu\,\tau}\,
  +2\,\ee\,\vtt \,
  e^{-\nu\,\tau}\,\cL_x(\tau)\,\cos\,\bOO\tau\,.
\end{eqnarray}
This result is similar to the one, obtained for guiding centers
correlation function~(see App.~\ref{sec:GC}).

\section{Complex number representation}
\label{app:xy_plane}

Making use of the homomorphism $\varphi$:
\[
        \varphi: A= 
        \left(
  \begin{array}[c]{rr}
    a & -b \\
    b& a
  \end{array}
\right) \to z=a+ib
\]
we can replace for the calculation in the $\{x,y\}$ plane:
\[
        \hat L_3=
        \left(
                \begin{array}[c]{cc}
                        0& -1 \\
                        1&  0
                \end{array}
        \right)\to i\,;
        \qquad R_3(\alpha)=
        \left(
        \begin{array}[c]{cc}
                \cos \alpha & -\sin\alpha \\
                \sin\alpha  & \cos\alpha
        \end{array}
        \right)
 \to e^{i\,\alpha}\,. 
\]
By the inverse transformation the diagonal elements of the matrix
are just a real part of the complex number\,:
\[z\to A=
\left(
\begin{array}[c]{cc}
        \Re [ z ] &     -\Im [z] \\
        \Im [z] &  \Re  [z]
\end{array}
\right)
\]

\section{Guiding Center Diffusion}
\label{sec:GC}

In this section we calculate the velocity correlation function
$\langle u(t_1)^i\,u(t_2)^j \rangle$ in a guiding center approximation
for particles, moving in a slightly perturbed homogeneous magnetic
field. In the limit of a strong magnetic field, a single particle
performs a spiral motion strictly along a magnetic field line. If the
magnetic field is unperturbed $(\bB_0= B_0\,\be_z)$, the velocity
correlator has a form:
\[ \langle u_0(t)^i\, u_0(0)^j \rangle  = 
\langle \bu_0(t)\otimes\bu_0(0)\rangle =
\vtt\,e^{-\nu\,t}\,e^{-\OO\,t\,\hL_3}\,; \] where
$e^{-\OO\,t\,L_3}=R_3(-\OO\,t)$ is a finite rotation matrix and $\OO$
is a Lamor frequency. The subscript~``0'' is used for the unperturbed
velocities.

Next we perturb the magnetic field, adding a small stochastic
component:
\[\bB(t)=B_0\,\be_z+ \e\,B_0 \,\vz(t)\,.\]
We suppose that the perturbation $\vz(t)$ yields the following Lagrange
correlation function:
\begin{eqnarray}
  \label{eq:app:LCF}
  \langle \vz(t)^i\,\vz(0)^j\rangle= 
  \left(
    \begin{array}[c]{ccc}
      \cL_x(t) & & \\
      & \cL_x(t) & \\
      && \cL_z(t)
    \end{array}
  \right).
\end{eqnarray}
Since the perturbation of the magnetic field is small, it can be
considered as a rotation of the unperturbed vector field $\bB_0$ at
each point of space (or time on Lagrange trajectory) to some
angle~$\alpha$ without change of its amplitude. This rotation can be
represented as
\[ \bB(t)= e^{\bk(t)\cdot \bL}\,\bB_0 \]
where $\bk$ is the rotation vector: $|\bk|=\alpha$ is the rotation
angle, and $\bn=\bk/k$ is the rotation axis. It is straightforward to
calculate (for $\e\to 0$):
\begin{eqnarray}
        \label{eq:k(t)}
        \bk(t)=\e\,[\vz(t)\times \be_z]\,.
\end{eqnarray}
Using the fact that in the strong magnetic field the guiding center
follows exactly the magnetic field line, we need not to calculate the
perturbed trajectory of the particle, rather we just have to apply the
same rotation operator $e^{\bk(t)\cdot \bL}$ to the unperturbed
velocities. The ``perturbed'' correlator becomes:
\begin{eqnarray}
  \langle u(t)^i\, u(0)^j \rangle = \langle e^{\bk(t)\cdot
    \bL}\,\bu_0(t)\otimes e^{\bk(0)\cdot \bL}\,\bu_0(0)\rangle
\end{eqnarray} 
Next we expand the rotation matrices, using the fact that $|\bk|\ll
1$\,:
\[
e^{\bk(t)\cdot\bL}\approx 1+\bk(t)\cdot\bL\,.
\] 
Substituting this expansion in the expression for the velocity
correlator, we obtain in index notation:
\begin{eqnarray}
  \label{eq:R}
  u(t)^i\, u(0)^j = u_0(t)^i\,u_0(0)^j+ [...] +
  k(t)^\alpha\,k(0)^\beta\,L_{\alpha\,l}^{\ \ i}\,L_{\beta\,m}^{\ \ 
    j}\,u_0(t)^l\,u_0(0)^m\,,
\end{eqnarray}
where $[...]$ implies terms proportional to the first order of
$\bk(t)$.

The vector $\bk(t)$ is a stochastic function, which correlators can be
readily found from Eqs.~(\ref{eq:app:LCF}) and (\ref{eq:k(t)}):
\begin{eqnarray}
  \label{eq:ck}
  &&\langle \bk(t) \rangle \equiv 0;\quad
  \langle k(t)^i\,k(0)^j \rangle= 
  \ee\,\cL_x(t)\,\Big(\delta^{ij}-n_z^{\,i}\,n_z^{\,j} \Big)\,,
\end{eqnarray}
where $n_z=\{0,0,1\}$.  Now we can average the expression (\ref{eq:R})
over magnetic fluctuations. Again we have to use the Corrsin
approximation, which justifies the independent averaging over
stochastic field (here over $\bk(t)$) and over the trajectory
variables (here over the unperturbed velocities~$\bu_0(t)$). In this
connection $\langle u_0(t)^i\,u_0(0)^j \rangle$ is the correlator on
unperturbed trajectory. Then, after straightforward but cumbersome
calculations, we get:
\begin{eqnarray}
  \label{eq:fR}
  \langle u(t)^i\, u(0)^j \rangle=\vtt\,e^{-\nu\,t}\,\Big( 
  R_3(-\OO\,t)+ 2\,\ee\,\cL_x(t)\,\cos(\OO\,t) + \ee\,\cL_x(t)
  \left(
    \begin{array}[c]{ccc}
      1&& \\
      &1& \\
      &&0
    \end{array}
  \right)
  \Big).
\end{eqnarray}
Comparison of this expression with Eqs.~(\ref{eq:hu35.1}) and
(\ref{eq:hu36.3}) shows that the $z$-component of the guiding center
correlator coincides exactly with that of particles. Yet the $\{x,y\}$
component of the particle correlator has an additional multiple
$e^{-\gamma(t)}$. In contrast to the particle diffusion, the last term
in the Eq.~(\ref{eq:fR}) gives a non-vanishing in the limit
$\OO\to\infty$ contribution to the guiding center diffusion
coefficient $D_x=\int_0^\infty \langle
u^x(\tau)\,u^x(0)\rangle\,d\tau$.

\bibliographystyle{unsrt}

\bibliography{bibfile}




\end{document}